\documentclass[conference]{IEEEtran}
\usepackage{adjustbox}
\usepackage{booktabs}
\usepackage[nospace,compress]{cite}
\usepackage{combelow}
\let\labelindent\relax
\usepackage{enumitem}
\setlist{leftmargin=1.2em}
\usepackage{float}
\usepackage[flushmargin,hang]{footmisc}
\usepackage{graphicx}
\usepackage{multirow}
\usepackage{stfloats}

\ifCLASSINFOpdf
\else
\fi
\usepackage[cmex10]{amsmath}
\newcommand{\MethodName}{DeepStamp}

\begin{document}
%
\title{\textbf{Poster:} On the Feasibility of Training Neural Networks with Visibly Watermarked Dataset
\vspace{-1.4em}}    

\author{
    \IEEEauthorblockN{Sanghyun Hong\IEEEauthorrefmark{1}, Tae-hoon Kim\IEEEauthorrefmark{2}, Tudor Dumitra\cb{s}\IEEEauthorrefmark{1}, and Jonghyun Choi\IEEEauthorrefmark{3}}
    \IEEEauthorblockA{\IEEEauthorrefmark{1}University of Maryland, College Park}
    \IEEEauthorblockA{\IEEEauthorrefmark{2}Deeping Source Inc.}
    \IEEEauthorblockA{\IEEEauthorrefmark{3}Gwangju Institute of Science and Technology (GIST)}
    \vspace{-2.0em}
}

\maketitle


\begin{abstract}
As there are increasing needs of sharing data for machine learning, there is growing attention for the owners of the data to claim the ownership.
Visible watermarking has been an effective way to claim the ownership of visual data, yet the visibly watermarked images are not regarded as a primary source for learning visual recognition models due to the lost visual information by in the watermark and the possibility of an attack to remove the watermarks.
To make the watermarked images better suited for machine learning with less risk of removal, we propose \MethodName, a watermarking framework that, given a watermarking image and a trained network for image classification, learns to synthesize a watermarked image that are human-perceptible, robust to removals, and able to be used as training images for classification with minimal accuracy loss. To achieve the goal, we employ the generative multi-adversarial network (GMAN). In experiments with CIFAR10, we show that the \MethodName\ learn to transform a watermark to be embedded in each image and the watermarked images can be used to train networks. 
\end{abstract}



\section{Introduction}
\label{sec:intro}

Convolutional neural networks (CNN) have achieved super-human performances in various computer vision and machine learning applications with the help of large-scale supervised datasets. 
However, the data curation is very expensive in terms of time, fund and much effort to control the quality of supervision.
To ease the cost of data curation, one can outsource the collection of a dataset to multiple parties with reasonable rewards.
However, there is a risk of the shared data could be stolen and shared without any rewards to the parties.
To prevent the case, we should be able to claim the ownership of the data when the data is stolen. 

One possible approach is to use cryptographic methods such as homomorphic encryption~\cite{gentry2009fully} and multi-party computation (MPC)~\cite{shokri2015privacy}. Homomorphic encryption allows computation on encrypted images and returns results when decrypted, matches the result of the operations as if they had been performed on the plain-data. MPC jointly computes a neural network over multiple inputs while keeping each input private. However, the computation on the encrypted data takes longer than that on the plain data. Also, once the content of the data is revealed, these schemes cannot protect the ownership since they are only designed to protect the privacy without sharing. Thus, they cannot prevent the fake ownership issues.

Another approach is to conceal secrets (messages) within data, when the data is shared with others, such as steganography or invisible watermarking~\cite{shih2017digital}. Unlike the cryptographic solutions, these methods do not introduce extra computational costs because they compute on the plain data. Nevertheless, the secrets are easy to be destroyed by the data modifications, e.g., cropping, rotating, or resizing, easily exploited by an attacker. Given that the modifications are commonly used in training CNNs for the data augmentation, if an attacker modifies the data before claiming the fake ownership, the owner is hard to prove whether the data is hers or not.  

We consider \emph{visible watermarking}~\cite{braudaway1996protect} to protect the ownership of shared data for readiness of information for claiming the ownership. There have been challenges in using the technique to the datasets for training CNNs because networks trained with watermarked data can suffer from accuracy losses, and an attacker can use sophisticated methods~\cite{dekel2017effectiveness} to remove watermarks from data. 
Our work first studies the accuracy drops to the changes in visibility and randomness\footnote{\cite{dekel2017effectiveness} shows that the random perturbation of watermark images embedded in data makes an attacker hard to remove the watermarks.} of the watermarks in data. We, then, propose \MethodName\ that synthesizes watermarks, once blended to data, minimizes the accuracy drop and makes them hard to be removed, with given data and a watermark image. We leverage a generative network, GMAN, to achieve randomness and implement necessary conditions as discriminators. With CIFAR10, we show that our watermarks minimize accuracy loss and, once we have watermarks, they can be used to train multiple CNNs.
 

\section{Threat Model}
\label{sec:threat}

We consider an adversary who \emph{claims the ownership} of the datasets produced or collected by others such as industrial partners or public sources. For instance, suppose that Alice wants to provide a data collection $A$ to Bob who wants to train CNNs using $A$ as a subset of their training data. However, Alice still wants to claim the ownership of the shared data $A$, to prevent the case that Bob turns into malicious and takes benefit from re-sharing/selling $A$ to other parties. Alice also can claim the ownership of $A$ when Bob did data modifications.

\section{\MethodName Framework}
\label{sec:method}



\begin{figure*}[t]
    \centering
    \includegraphics[width=0.972\textwidth]{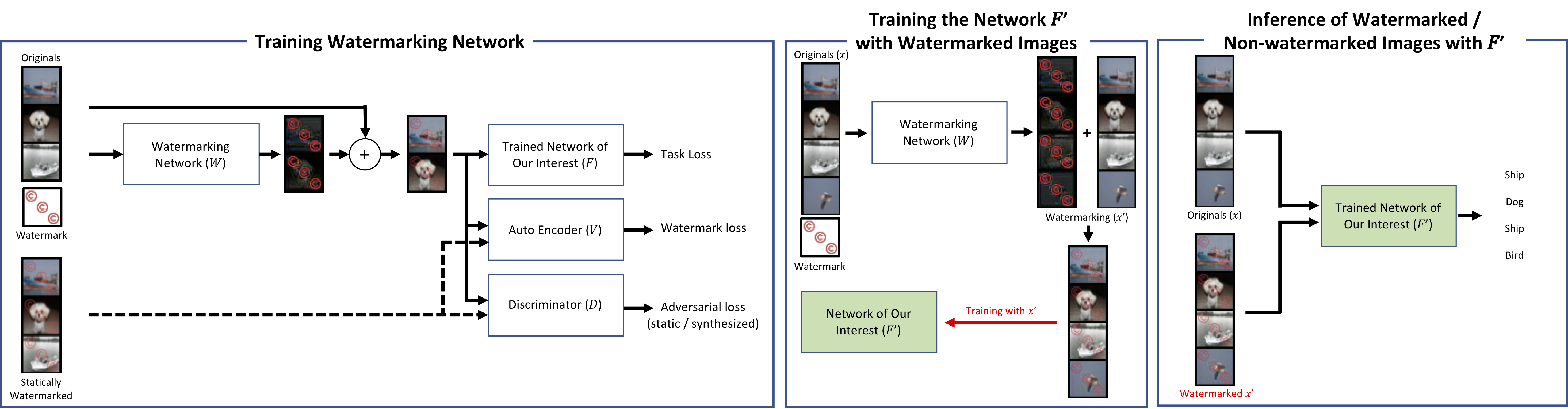}
    \caption{\textbf{Overview of Our Watermarking Framework.} We employ Generative Multi-Adversarial Networks (GMAN) to train the watermarking netwotk ($W$) which returns task-aware watermark images $w$ for each corresponding original data ($x$). We embed $w$ to $x$ and train the network of our interest ($F'$) with the watermarked images ($x'$). Once the network is trained, the network ($F'$) infers the correct labels for both $x$ and $x'$.}
    \label{fig:watermarking-framework}
    \vspace{-1.4em}
\end{figure*}

We aims to learn transformations of a given watermark to each data that can: (1) minimize the accuracy drop of networks trained on the watermarked data, (2) make the watermarks embedded in each data hard to remove by an attacker, and (3) make watermarks clearly perceptible to human-eyes. 
We illustrate the overview of our framework in Fig.~\ref{fig:watermarking-framework}. First, since \cite{dekel2017effectiveness} revisits that the random perturbation of watermarks embedded to data makes the watermark robust to removals, we learn the random perturbation for each data automatically during training by employing a generative network, GMAN, that enables a watermarking network $W$. To minimize the accuracy loss, we utilize a pre-trained CNN model $F$. We use an auto-encoder $V$ and a discriminator $D$ to enforce the transformations from $W$ are visually similar to the original watermark at a certain level. 

\noindent \textbf{In training}, \MethodName\ minimizes:

\begin{itemize}[topsep=0em,itemsep=0.2em,partopsep=0em,parsep=0em]
    \item $\mathcal{L}_{f}(x, x_w')$: the task loss computes the difference in the inferred labels of the clean data $x$ and the same data with synthesized watermarks $x_w'$.
    \item $\mathcal{L}_{v}(w, w')$: the $\ell_2$ loss from the auto-encoder $V$ between the original watermark $w$ and the synthesized one $w'$.
    \item $\mathcal{L}_{d}(x_w, x_w')$: the discriminator loss (binary cross entropy) that separates the original watermarked data $x_w$ from the data with synthesized watermarks $x_w'$.
\end{itemize}

\noindent Thus, the entire loss that we minimize is:

\setlength{\abovedisplayskip}{-1.0em}
\setlength{\belowdisplayskip}{0.2em}

\begin{gather}
    \vspace{-1.0em}   
    \mathcal{L}_{tot} = \mathcal{L}_{f}(x, x_w') + \mathcal{L}_{v}(w, w') + \mathcal{L}_{d}(x_w, x_w').
\end{gather}

\noindent \textbf{In stamping}, \MethodName\ synthesizes the watermarked data $x_w'$ for a given data $x$ and an watermark image $w$ that is visible and hard to remove with minimal accuracy drop. Alice now can share the watermarked data $x_w'$ to Bob, and Bob can train his network $F'$ with $x_w'$. Alice does not concern about the ownership issue since $w'$ is visible, and $w'$ is hard to be removed by anyone else because each $w'$ has randomly perturbed by $W$. Once $F'$ trained, $F'$ ensures the similar accuracy on both the $x$ and watermarked data $x_w'$.

\section{Evaluation}
\label{sec:evaluation}


\begin{table}[b]
\begin{adjustbox}{width=\linewidth}
    \begin{tabular}{@{}ccccccl@{}}
    \toprule
    \multirow{2}{*}{\textbf{Network Arch.}} & \textbf{CIFAR10} & \multicolumn{5}{c}{\textbf{Watermarked CIFAR10}} \\ \cmidrule(l){2-7} 
     & \textbf{Baseline} & \textbf{Blend} & \textbf{S} & \textbf{O} & \textbf{D} & \textbf{\MethodName} \\ \midrule \midrule
    \multirow{2}{*}{\textbf{AlexNet}} & \multirow{2}{*}{82.74} & 0.5 & 78.50 & 79.10 & 80.13 & \multicolumn{1}{c}{79.59} \\
     &  & 1.0 & \multicolumn{1}{c}{73.51} & \multicolumn{1}{c}{73.15} & \multicolumn{1}{c}{73.62} & \multicolumn{1}{c}{74.09} \\ \midrule
    \multirow{2}{*}{\textbf{VGG16}} & \multirow{2}{*}{94.00} & 0.5 & 92.71 & 92.92 & 92.58 & \multicolumn{1}{c}{92.74} \\
     &  & 1.0 & \multicolumn{1}{c}{92.57} & \multicolumn{1}{c}{92.83} & \multicolumn{1}{c}{92.61} & \multicolumn{1}{c}{-} \\ \midrule
    \multirow{2}{*}{\textbf{ResNet50}} & \multirow{2}{*}{95.37} & 0.5 & 94.88 & 94.71 & 94.92 & \multicolumn{1}{c}{94.18} \\
     &  & 1.0 & \multicolumn{1}{c}{94.67} & \multicolumn{1}{c}{93.64} & \multicolumn{1}{c}{93.66} & \multicolumn{1}{c}{-} \\ \bottomrule
    \end{tabular}
\end{adjustbox}
{
    \begin{flushright}
        \scriptsize (\textbf{S}, \textbf{O}, and \textbf{D} indicate the \textbf{S}tatic method, \textbf{O}pacity variation, and \textbf{D}isplacement in~\cite{dekel2017effectiveness}.)
    \end{flushright}
}
\caption{Accuracy of Models Trained with W'marked Data.}
\label{tbl:acc-drops}
\end{table}

\noindent \textbf{Experimental Setup.} We use CIFAR10 that consists of 32x32 pixels, three channel images. The images are labeled into ten classes, containing 50,000 training and 10,000 validation images. We implement the network $W$ with four convolutional layers whose input is concatenation of three channel data and four channel watermark (so total of seven channel) and output is a four channeled watermark. $D$ is composed of three transposed convolutional layers, and $V$ has five convolutional layers. For the network of our interest $F,F'$, we use three popular CNN architectures: AlexNet, VGG16, and ResNet50.

\noindent \textbf{Results.} We summarize the results in Table~\ref{tbl:acc-drops}. We observe the followings:
\begin{itemize}[topsep=0em,itemsep=0.2em,partopsep=0em,parsep=0em]
    \item Visible watermarking causes the accuracy drops in all cases, however, if the network capacity (the number of parameters in a network) is higher, the acc. drop is lower.
    \item When we use a strong blending factor (1.0), the accuracy drop increases. However, the drops are not significant with a high-capacity network (ResNet50).
    \item Using AlexNet for $F,F'$, our data with synthesized watermark (\MethodName) has a less accuracy drop than the statically watermarked data (S). However, we are not better when the watermarked data with displacements (D) is used.
    \item By training VGG16 or ResNet50 ($F'$) with our data synthesized with AlexNet ($F$), we observe accuracy drops by 1.17\% and 0.74\% compared to the static (S) method. Since the drops are similar to the case in which we use the same networks as $F$, once synthesized, the data can be used to train multiple $F'$s.
\end{itemize}

\section{Conclusion}
\label{sec:conclusion}

We propose a watermarking framework, \MethodName, that embeds a visible watermark into the images of interest with less accuracy drop and difficulty of removal for easy claim of the ownership of the watermarked images in the data-sharing scenario. 
In experiments with the CIFAR10 dataset, we show that the \MethodName\ learns transformations of a watermark to be embedded in another images with negligible accuracy drop while making its removal from the images non-trivial.


\section*{Acknowledgment}

This research is partially supported by Department of Defense and the ``Global University Project" grant funded by the Gwangju Institute of Science Technology (GIST) in 2018.


\bibliographystyle{IEEEtranS}
\bibliography{bib/security}

\end{document}